\documentclass{article}
\usepackage{amsmath,amsfonts,amssymb,mathrsfs,graphicx}
\begin{document}
\title{Generalized Entropy from Mixing: Thermodynamics, Mutual Information and Symmetry Breaking}
\author{Fariel Shafee\footnote{This invited review paper was written in January 2009 based on three papers within the author's Princeton University thesis. The author has now left Princeton}\\
Physics Department, Princeton University, Princeton NJ 08540, USA\\
current E-mail: fshafee@alum.mit.edu}

\maketitle

%$^{\star}$Author to whom correspondence should be addressed.\\[12pt]
%{\em Received:  / Accepted:  / Published: }}

% Abstract
\begin{abstract}
We review a new form of entropy suggested by us, with origin in mixing of states of systems due to interactions and deformations of phase cells. It is demonstrated that this nonextensive form also
leads to asymmetric maximal entropy configurations unlike Shannon entropy. We discuss how, beginning with
quantum entanglement of microsystems with one another and with the environment, one can obtain classical
stochasticity for our form of entropy.

\vspace*{0.5cm}

\end{abstract}
%==========================================================

%==========================================================
% Main Text of the Paper
%----------------------------------------------------------

%%%%%%%%%%%%%%%%%%%%%%%%%%%%%%%%%%%%%%%%%%%%%%%%%%%%%%%%%%%%
\section{Introduction}

Entropy is a measure of disorder or uncertainty in a system.  The most traditional form of entropy, Boltzmann/Gibbs, was proposed to understand ensembles where certain macro variables, and not the exact state of each unit, are known, so that a fuzzy collection of microstates represent the same macro-state.  Statistical distributions of microstates are derived by using relations of the entropy of a system with the measurable averaged properties of the macro-state such as temperature.  The greater the number of possible microstates a system occupies (i.e., the larger the number of possible states the ensemble is distributed over), given a certain set of macro-variables, the larger the entropy of the corresponding macro-state.  In information theory, the concept of entropy is related to the uncertainty of obtaining a particular variable in a series of possible combinatorial outcomes, and the most well-known information entropy is given by Shannon's form.  If the variable can be predicted with certainty, the system measured has zero entropy.

Novel definitions of entropy were proposed in recent years based on various concepts of order and uncertainty related to different types of systems and interactions among the parts of the systems \cite{VN1, Kol1, Kol2, TS1}.  Some of these forms \cite{Kol1, Kol2} describe the degree of order and predictability in the trajectory of dynamic systems, and some \cite{TS1} take into account the constraints imposed by the interactions among the parts of the system.  Although, in the Boltzmann/Gibb's form of entropy, the units in the ensemble are taken to be independent of one another's influence, so that every possible microstate is equally likely to occur, as if in isolation, many physical phenomena showing significant inter-unit correlations \cite{SUY1, GRA1} are better described by newer forms of entropy, such as Tsallis's \cite{TS1}.

This trend of including more complex physical behavior into the concept of entropy was furthered by another form of entropy \cite{FS1} proposed by us. The new form of mixing entropy takes concepts of order and predictability associated with complexity and information in the real world into account.

While the traditional and most other known forms of entropy assume the number of possible observable states to be pre-defined, so that the uncertainties and the degrees of disorder in the system originate from the possible distribution of these well-defined states within a closed system, the notion of states and identities in the environment is much more complicated, and often evolve due to the interactions between the units of the systems or the units and the larger environment.  New forms of identity arise as new possible states because of these interactions between different forms of modified and extended identities at different hierarchical levels within the environment.

Although each expressed identity in the environment can be broken down into consequently smaller fragments, reducing to elementary particles in the end \cite{GM1,SM1,FS8}, it is impossible to find a statistical distribution of the universe based on a Boltzmann-Gibbs-like probability distribution of elementary particles. The various forces of nature causing the elementary particles to form clusters and possible repetitive patterns of organization with different degrees of stability are not taken into account in traditional statistical distributions.  In the macroscopic world, somewhat stable and repeating correlated systems are assigned identifications and hence states, characterizing bundled properties, and relatively similar systems are categorized \cite{categor}.  For example, it is possible to isolate bacteria based on some general characteristics displayed by the correlated molecules within all strains of bacteria, and bacteria can be further divided into possible strains. The distribution of such strains can be studied in a system consisting of bacteria even though these bacteria are not completely isolated from the larger environment. Such systems are neither closed nor completely stable.  It can be said they are somewhat semi-closed \cite{FS8} and semi-stable, displaying some known characteristics based on past knowledge, while also allowing for changes because of the interactions among themselves and due to interactions with many degrees of the larger environment.  However, it may still be possible to derive statistical properties of these complex systems because of the nature of their semi-stability, accounting for possible changes of identities due to the mixing of the interacting atoms allowing for new orders of correlations among such categorized states.

The new form of entropy proposed is based on the concept of mixing of information in the identity containing cells of the
system.  The entropy yields a non-extensive form, because of interaction among the units of the ensemble. The derivation of the entropy is from first principles, using information theoretic measures and taking into account the number of different states and the added degrees of freedom introduced in the  deformed cells on account of mixing.

We then review the derivation of the mixing entropy and the thermodynamical properties related to this generalized quantity.  The differences of the mixing entropy with existing forms of entropy are then pointed out. Possible practical applications of this new disorder measure are also suggested. An interesting property similar to spontaneous symmetry breaking of field theory that can be associated with this entropy form is elaborated. Lastly, the classical definition of this mixing entropy is extended to quantum information contexts, and the effect of mixing information due to deformation of cells in the calculation of mutual information is examined. We also briefly review the origin of stochasticity by taking into account the entanglement with the environment, and then apply the new form of entropy to such systems.

Towards the end of the paper, the philosophical implications of the new form of entropy and its in-depth origin related to complex systems and leakage of information are discussed.  The terms used in defining the entropy such as `cells', `letters' and ``deformation'', are also explained in detail in the context of this specific entropy and information in complex systems.  The last section clarifies the differences between the origin and application of the new form of mixing entropy and other existing forms of entropy.

\section{Units of Mixing}
In the rest of the paper, we will be discussing mixing of letters of an alphabet as a consequence of cell deformation.  Here, we briefly define the terms.  The philosophical implications of these terms with examples and their possible applications are discussed towards the end of the paper.

\subsection{cells, identities and deformation}
A cell or registrar holds information that can be expressed together as a unit identity.  The information contained within the cells manifest as specific states of a variable. Hence, a cell contains a variable that can be measured to give a state of the variable. Each cell is expressed as a unit of expression so that the entire information content of a cell indicates a certain state of the cell.  In a computer, a registrar may contain a bit which can be either in a 0 state or in a 1 state.

An alphabet is a known sequence of patterns. These patterns indicate possible states so that a certain state is tagged with a letter.  Recognising a pattern (or letter) indicates measuring a specific known state.  The alphabet encompasses the set of all known states.
A binary alphabet contains only two letters: 0 and 1.

Hence, an alphabet consisting of N letters indicate N possible outcomes or states of a measurement of a cell.  These N letters are known by previous experience and can be assigned a probability distribution, which gives the expected frequency of each of the letters in a series of measurements.

Let us assume that a cell can hold enough information to indicate only one letter and the letters are all predefined  as possible states of a variable by previous knowledge.  Hence, past experience would assign an alphabet consisting of all possible outcomes of measuring a cell.  The probabilities of each of the letters appearing as a result of measuring a cell would then add up to one, and the expected states are assumed to be known by past knowledge.

A cell holding a letter yielded by such a measurement can be scaled to a phase space or volume of one.  Hence, the scale of a cell is defined such that each such registrar scaled to ``one'' is allowed to hold exactly one letter or known state.

However, these cells are part of a larger environment, and the states are expressed within a larger dynamical hierarchical structure of the world. Hence, the cells containing possible states of a single variable also interact with the environment and other variables.  As a result, the phase space or volume of each cell may change.  This change indicates a unit of measurement or a unit of identity being allowed to hold more or less information than it did before.

A simple example of distorting information containing phase space would be to inflate a balloon so that the volume inside the balloon is changed. The balloon can hold more matter within it and the newly introduced matter may get mixed with the matter previously contained within the balloon.  If the balloon was a cell, the inflated balloon is a deformed cell that allows more information inside it.

\begin{figure}[ht!]
\begin{center}
\includegraphics[width=10cm]{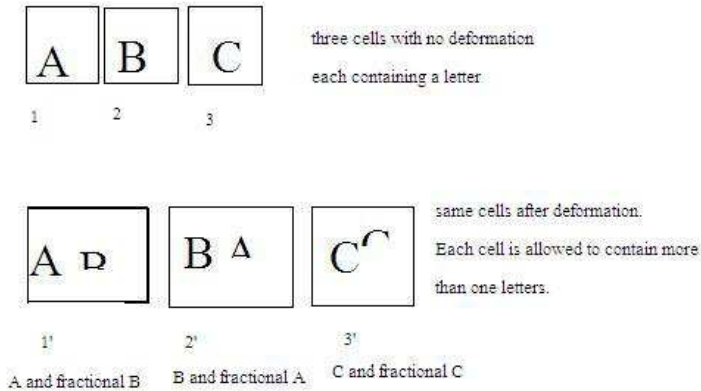}
\end{center}
\caption{\label{fig5025} Information content of cells before and after deformation. With no deformation, each cell has adequate space to hold information to define one letter.  After deformation by a factor q, each cell can hold more information, and hence may have mixed fractional letters.}
\end{figure}

\begin{figure}[ht!]
\begin{center}
\includegraphics[width=10cm]{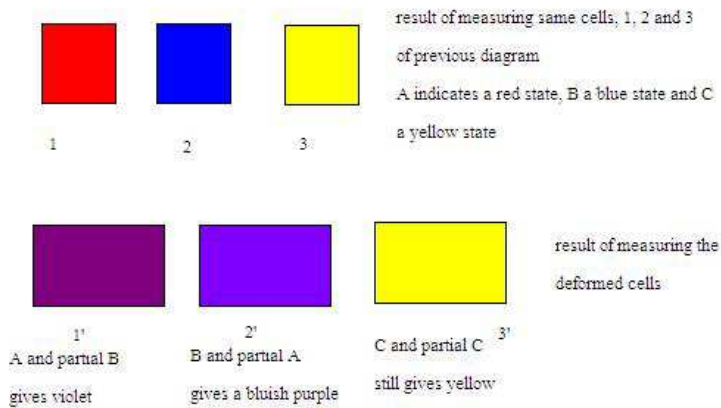}
\end{center}
\caption{\label{fig5025} The results of measuring the cells defined in the previous diagram.  If each letter in the cells represents a certain color, the results after deformation may yield new colors after mixing of states.  The amount of mixing will determine the exact shade.}
\end{figure}

If the original cell was big enough to contain information to be expressed as one letter or state, a deformed cell can now hold a different amount of information so that it can hold information in excess of what is needed to define a previously known state.  Particularly, it can have information related to another state mixed with the original state of the registrar.  If the original state of the registrar indicated a pattern that could be recognized as a letter, the deformed cell can now hold that previous letter and some fragments of another letter or state ``flowing into it'' from the environment or other cells because of the changed capacity of the registrar.

We say that the original set of letters defining the alphabets, indicating previously known states of the measured variable are pure letters.  After a cell or registrar is deformed so that its information containing phase space is changed, the cell is q-deformed, where q is the measure of excess (or less) information content allowed by the cell.  The new content of a q-deformed cell may be mixed letters, where partial information of different letters are entangled within a measurable cell, making the variable state different from the previously known letters.  The degree of mixture and the letters within the mixture will may indicate new properties associated with the measured mixed state.

A more in depth explanation of the process and the philosophical implication with respect to complexity theory and expressed identities in a complex environment where possibly distorted variables may exist is towards the end of the paper.  The exact meaning of the deformation variable q, and its difference from the ``q''s in other forms of entropies are explained in detail in the last part of the paper.

\section{Defining the New entropy}

\subsection{Non-extensivity in Entropy}

Nonextensivity of entropy is a field of current interest because of its relation with complexity and correlated systems.  While for extensive forms of entropy, one considers subsystems of a larger system to be independent of each other, forming  statistical distribution based on laws of probability while macroscopic parameters like energy and temperature of the system come into play as constraints to preserve the {\em average} properties of the ensemble, correlations among units are not negligible in specific complex organizations \cite{WE1, LI2}.  Tsallis \cite{TS1} proposed the first important form of non-extensive entropy, using the idea of generalized logarithms, and a parameter, $q$, which is the measure of deviation from extensivity.  This nonextensive form of entropy was able to phenomenologically explain correlated events like biased risk taking in gambling \cite{SUY1} and dynamic correlated systems \cite{GRA1} and other situations. The non-extensivity of this form of entropy is derived from the correlations of sub-systems within the system itself, so that the measurement of one variable depends on the presence of another variable, making the total disorder of the system deviate from the sums of the separate disorders of the subsystems.  The idea of Tsallis entropy \cite{TS1} was derived from the existence of powerlaw behavior in the probability distributions of many natural phenomena including fractals \cite{LY1}.  The logarithm of Boltzmann-Gibbs entropy was generalized to include a parameter q that empirically indicates the deviation from extensivity within a system, and hence correlation of states within a certain system.

The mixing entropy reviewed here is also non-extensive. However, the non-extensivity of this new form of entropy derives from the leakage of information and the degree of permeability between states produced by deformation of information containing registers.  The interaction between the system and the measurer is also taken into account.  While the possibly observable states are kept constant in most entropic measurements, this new form of entropy takes into account the evolution of states themselves by allowing information indicating states to mix by leakage, hence allowing a known pure state to evolve into a new mixed state, which acts as a new unit. If the possible states of measurements are indicated by letters and the cells containing the letters are allowed to deform, then the level of purity maintained within each cell is dependent on the leakage of information from a different letter contained within another cell.  The leakage of the same letter into a deform cell does not add any extra information, and hence does not create a new identity even if the cell is deformed in our definition.  A detailed mechanism for the origin of the mixing entropy is discussed towards the end of the paper.

While a criticism of Tsallis entropy \cite{ZA1} is that basically any statistical distribution can be obtained by changing the function whose generalized mean is taken as a constraint, the probability distribution function (PDF) of this new entropy depends on the properties of the Lambert function, which is fixed by the nature of this entropy.

\subsection{The Mathematics of Mixing}

The new form of mixing entropy \cite{FS1} may be derived using information theory concepts. We outline the approach below:

An alphabet of $N$ defined letters is considered.
Originally, registers are given the capacity to hold single letters and $p_i$ is the set of probabilities assigned for each of these $N$ letters, $A_i$, to appear in one of these registers or cells. This information-theoretic description can be trivially extended to states $i$ of a single microsystem in an ensemble. Each of the individual subsystems can then be in any one of $N$ states with probabilities $p_i$.

If a small distortion is introduced so that each of these single cells is allowed to contain $q = 1+ \Delta q$ letters, the new probability that the entire deformed phase space is occupied by the letter previously associated with $p_i$ becomes $p_i^q$.  The deformed cell now contains any single pure letter form in the alphabet, $A_i$,  with the probability:

\begin{equation} \label{eq501}
N(q) =  \sum_i p_i^q
\end{equation}

The original total probability that a register contains a pure letter is unity for the case $q=1$.  If $q >1 $ the total probability that the register contains a pure letter becomes less than 1.  The difference is denoted by

\begin{equation} \label{eq502}
M(q) = 1- \sum_i p_i^q
\end{equation}
This is a measure of mixture between $A_i$ and other members of the alphabet ($A_j; j \neq i$) since the total probability that the cell is occupied must be $1$. The introduction of possible fractional mixed letters thus defines a quantity,  $M(q)$, which expresses the degree of disorder created by changing the cell scale to $1+ \Delta q$.

%%%%%%redo very very crefully: renyi already takes fractional dimensions into account

%%%%%%%%%%%%%%%%%%%%%%%%%%%%%%%%%%%%%%%%%%%%%%%%%%%%%%%%%%%%%%%%%%%%%%%%%%%%%%
%%%%%%%%%%%%%%%%%%%%%%%%%%%%%%%%%%%%%%%%%%%%%%%%%%%%%%%%%%%%%%%%%%%%%%%%%%%%%%
\subsection{The Origin of Uncertainty in Mixing}

The origin of the new entropy can be understood in part by looking into the definition of entropy in the context of information theory.
In information theory, entropy is a measure of uncertainty of states of a typical system in an ensemble, as identified by differences of at least one variable.  When measurements are performed, deformation of cells can introduce uncertainties in the identities of the detectable letters as described above.  This new degree of uncertainty can be associated with change of the scale and the modification of the ability to retrieve a pure state associated with the original scale of measurement.

Consequently, the effective state space expands to accommodate probabilistic {\em mixed} letters not present previously. The original alphabet consists of $N$ letters, and hence gave rise to an $N$-dimensional state space volume. The addition of mixed states gives new contributions to the uncertainty.
An in depth analysis of the origin of such uncertainty in interacting systems can be found towards the end of the paper.

This new entropy can be given an entropy density form, $S(q)$, as a function of the exact scaling in the same spirit of Kolmogorov-Sinai entropy densities \cite{KOL1}. The net possibility of mixing introduced by this change in scale is $M(q + \Delta q) - M(q)$.

Let us assume that an alphabet has N letters.  Using usual definitions from information theory, and also the effect of mixing introducing more information content within a registrar, the information held by a registrar can be given by

\begin{equation}
N^{S(q)\Delta q} = N^{M(q+\Delta q)-M(q)}
\end{equation}

S(q) is the entropy density given a deformation of the cell by q. $S(q) \Delta q$ yields the total measure of uncertainty introduced into the registrar by expanding it from the scale $1+q$ to $1+q+\Delta q$.  The uncertainty in the information content, denoted by this entropy, responsible for increased content in the registrar and hence the presence of more than one letters, is  introduced by mixing, and is given by the change in mixing probability between the scaled $1+q$ and $1+ q + \Delta q$.

From this relation, the definition of the new entropy density, $S(q)$ can be found as

\begin{equation} \label{eq504}
S(q)  = dM(q)/dq
\end{equation}
or, equivalently,

\begin{equation} \label{eq505}
S(q) = - \sum_i p_i^q \log p_i
\end{equation}

This is the mathematical definition of our form of entropy, and signifies the rate of the the possibility of mixing with the rate of changing cell size at a certain scale, q. The ability of a cell to contain a pure letter even if the cell is slightly deformed is can be associated with a degree of uncertainty associated with the system at that deformation scale, q.

%%%%%%%%%%%%%%%%%%%%%%%%%%%%%%%%%%%%%%%%%%%%%%%%%%%%%%%%%%%%%%%%%%%%%%%%%%%%%%%%%%%%%%%%%%%%%%%%%
%%% start from here

\subsection{Consequences of Cell Deformation in Shannon's Coding Theorem}

%%%new
We consider the simple case of information mixing within a series of letters.  The redefined distribution of pure letters allowing for leakage between information contents of cells exhibits a form of non-extensivity and modified entropy.  The ergodic nature of the series is slowly lost as the scale of measurement or information content is slowly allowed to vary with time, giving rise to a new type of dynamic system.

Shannon coding theorem \cite{SH1} states that in the infinite time limit, the frequency of the letters, $A_i$, appearing in a sequence determines the entropy density of the sequence, and hence the probability of a particular sequence appearing.  So, if the sequence has $n$ letters, and the frequency of each alphabet being expressed is $n p_i$, and the probability of a typical sequence

\begin{equation}
P(sequence) = \prod_i p_i^{n p_i}= \exp[-n S]
\end{equation}
where $S$ is the entropy per unit.
If the scale is changed so that a letter is measured in a $q$-deformed register, the frequency of a pure decipherable letter in the sequence also changes, and the new equation is
\begin{equation}
P(sequence) = \prod_i p_i^{n p_i^q}= \exp[-n S_{new}]
\end{equation}

So,
\begin{equation}
S_{new}= -\sum_i p_i^q \log p_i
\end{equation}

%%% new

Hence, the possibility of mixing of information in each scale because of deformation of cell, yielding an uncertainty in the retention of pure letter for each cell, manifests itself as a modified entropy density for a sequence.  The density of entropy in the distribution of pure previously defined letters (at unit scale)after the possibility of mixing is taken into account for each cell, is expressed by $S_{new}$, which has the same expression as the same uncertainty each cell experiences in retaining a pure letter after the deformation.

%%%%%%%%%%%%%%%%%%%%%%%%%%%%%%%%%%%%%%%%%%%%%%%%%%%%%%%%%%%%%%%%%%%%%%%%%%%%%%

%%%%%%%%%%%%%%%%%%%%%%%%%%%%%%%%%%%%%%%%%%%%%%%%%%%%%%%%%%%%%%%%%%%%%%%%%%%%%%%%%%%%%%%%%%%%%%%%%%%%%%%
%%%%%%%%%%%%%%%%%%%%%%%%%%%%%%%%%%%%%%%%%%%%%%%%%%%%%%%%%%%%%%%%%%%%%%%%%%%%%%%%%%%%%%%%%%%%%%%%%
%%%%%%%copied from next paper

%%%%%%%%%%%%%%%%%%%%%%%%%%%%%%%%%%%%%%%%%%%%%%%%%%%%%%%%%%%%%%%%%%%%%%%%%%%%%%%%%%%%%%%%%%%%%%%%%%%%%%%%%%%%%%%%%%%%%%%
%%%%%%%%%%%%%%%%%%%%%%%%%%%%%%%%%%%%%%%%%%%%%%%%%%%%%%%%%%%%%%%%%%%%%%%%%%%%%%%%%%%%%%%%%%%%%%%%%%%%%%%%%%%%%%%%%%%%%%%

\section{A Specific Thermodynamic Example in Ecosystems }
%semi-closed systems and perturbation
%the effect of one on another and deformation
In \cite{FS8}, the concept of semi-closed structures was discussed.  These are physical complex units that are almost closed with respect to their identity, but may get distorted because of interactions with the environment and other such units.  The complexity of these organizations accommodates many unit building blocks organized within a hierarchical structure. The level of possible distortion of these semi-closed systems depends on the interactions among them, and also the robustness of each structure's internal coupling. The leakage of information in such structures may be due to interactions among units organized at different hierarchical levels, and expressed as parameters defining the identities of these semi-closed units.

For example, let us consider two strains of bacteria that can exist in macroscopic colonies within a habitat.  Let the optimal size of each colony be fixed by environmental properties such as need for symbiosis, density of food and pH level of the environment. Then the number of colonies of each strain gives a measure of the frequencies or probabilities of each type of bacteria. The optimal size of a colony can be changed because of readjusted pH conditions, and the readjustment of colony size may cause bacteria to migrate from one colony to another causing mixing between two strains, and impure clusters are created.

The mixing origin of entropy may be relevant to thermodynamic systems such as clusters of molecules and living organisms etc., where the presence of other such clusters may distort the identities of one another, and the interactions among the clusters, and the energetic stability of sub-systems of such systems can also dictate the probabilities of the building blocks to readjust given the constraints imposed by the macroscopic environment.

\section{Probability Distribution from Thermodynamic Constraints}

Remarkably, the probability distribution function corresponding to this new entropy was found in a closed form \cite{FS1} by maximising the mixing entropy, given a value of  the deformation parameter $q$.

We start by assigning specific energy values to the probability states, and making the total energy of the system a constant.

We have the equation for the constrained optimization function $L$ , with the Lagrange multipliers $\alpha$ and $\beta$,

\begin{equation}\label{eq5015a}
L = S + \beta ( \sum_i p_i E_i - U) + \alpha( \sum_i p_i -1)
\end{equation}

Hence, when $S$ is the new form of entropy, optimization in $p_i$ gives

\begin{equation}\label{eq5015b}
- \frac{q}{q-1} p_i^{-(q-1)} (  \log p_i - 1)+ 1 + \gamma
p_i^{-(q-1)} = 0
\end{equation}
when optimized with respect to the individual probabilities.

Here, $\gamma= \alpha+ \beta E_i$ for clarity.

Some algebra gives \cite{FS1},

\begin{equation}\label{eq5016}
p_i = \left[\frac{ - q W(z)}{(\alpha+ \beta E_i)(q-1)}\right]^{1/(1-q)}
\end{equation}
where
\begin{equation}\label{eq5017}
z=-e^{(q-1)/q} (\alpha+ \beta E_i) (q-1)/q
\end{equation}
and $W(z)$ is the Lambert function defined by \cite{VA1}

\begin{equation}\label{eq5018}
z= W(z) e^{W(z)}
\end{equation}

When $q \rightarrow 1$, the original Gibb's distribution is retrieved.

For the new entropy, the non-extensive property can be found as

\begin{equation} \label{eq5023}
S{}_{1+2} = S{}_1 + S{}_2 - M_2(q)S{}_1 - M_1(q)S{}_2
\end{equation}
where the $M_a$ are the mixing probability of states for subsystem
$a$ as defined in Eqn.~\ref{eq502}.

A detailed analysis of the PDF with respect to the branches of the Lambert function for various cutoff of energies is shown in \cite{FS1}.

%%%%%%%%%%%%%%%%%%%%%%%%%%%%%%%%%%%%%%%%%%%%%%%%%%%%%%%%%%%%%%%%%%%%%%%%%%%%%%%%%%%%%%%%%%%%%%%%%%%%%%%%%%%%%%%%%%%%%%%

%%%%%%%%%%%%%%%%%%%%%%%%%%%%%%%%%%%%%%%%%%%%%%%%%%%%%%%%%%%%%%%%%%%%%%%%%%%%%%%%%%%%%%%%%%%%%%%%%%%%%%%%%%%%%%%%%%%%%%%

The PDFs for the new mixing entropy and for Boltzmann-Gibbs (B-G) distribution are given in Fig.~\ref{fig5025} for different values of $E$ and $q$. The black line is for $q=1$, corresponding to the Boltzmann-Gibbs case.

\begin{figure}[ht!]
\begin{center}
\includegraphics[width=10cm]{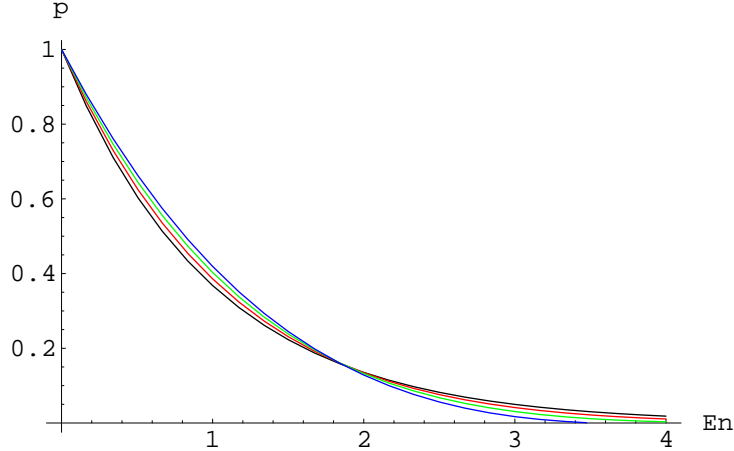}
\end{center}
\caption{\label{fig5025} The PDFs of the new entropy for $q=1,1.05,1.10$ and $1.15$.  The lines are in the order of $q$, with black line for $q=1$.}
\end{figure}

The deviation from the Gibbs distribution is discernible for large values of $q$.

\section{Thermodynamic Properties}

For any entropy function, a function $\phi(p_i)$ can be defined for each of the possible states so that the entropy is a sum of all these component functions:

\begin{equation}\label{eq5042}
S = \sum_i \phi(p_i)
\end{equation}

After optimizing the constraint equation, and carrying out some algebra, the PDF can be obtained as
\begin{equation} \label{eq5058}
p_i = \psi^{-1} ( \beta(E_i- A))
\end{equation}
where

\begin{equation}\label{eq5059}
\psi(p) = \phi(p)/p
\end{equation}

$A$, the free energy, can be calculated by using the equation and also the relation

$\beta (U- A)= S$, as

\begin{eqnarray} \label{eq5060}
\sum_i p_i = \sum_i \psi^{-1} (\beta (E_i -A)) = 1
\end{eqnarray}

The specific heat, $C$, can also be found by using

\begin{equation}\label{eq5061}
C= -\beta^2 \frac{\partial U}{ \partial \beta}
\end{equation}

\section{ Comparison with B-G and Tsallis Thermodynamics}

Using Eqn.~\ref{eq5058} and Eqn.~\ref{eq5059}, the B-G statistics yields

\begin{eqnarray}\label{eq5062}
p_i = e^{-\alpha - \beta E_i}= e^{\beta(A- E_i)}
\end{eqnarray}

Both the new mixing entropy and Tsallis entropy approach this distribution in the first order expansion of their respective functions (W function for the mixing entropy and generalized log for Tsallis).

At the second order the new mixing entropy and Tsallis form can be related so that the PDF for the new entropy is
\begin{equation}
e^{ -\log \left(1+ \epsilon_T \beta(E_i -
A)\right)/\epsilon_T}
\end{equation}

$\epsilon_T$ is the Tsallis parameter.  \cite{FS1} gives detailed calculations.

In the Gibb's case Eqn.~\ref{eq5060} yields, for $A$

\begin{equation}\label{eq5063}
A = - \log(Q)/\beta
\end{equation}
where $Q$ is the partition function

\begin{equation}\label{eq5064}
Q = \sum_i e^{-\beta E_i}
\end{equation}

In the case of B-G entropy, A can be calculated easily, giving us extensive thermodynamics.

For the Tsallis case, however, $A$ cannot be found with respect to the partition function. Instead, if
Eqn.~\ref{eq5060} is solved, an
infinite number of roots  appear for $\epsilon \equiv q-1$, except for the specific instances when $\epsilon$ is the inverse of an integer.  For those cases, a finite number of roots can be found.

The specific heat for the new mixing entropy can be found as

\begin{equation} \label{eq5073}
C = \beta^2 \sum_i  E_i (E_i-A)\frac{e^{-W^i( 1+1/\epsilon)}}{1+ W^i}
\end{equation}.
Here $W^i = W_0 \left(\epsilon \beta(E_i-A)\right)$
Details of the calculation can be found in \cite{FS1}

\begin{figure}[ht!]
\begin{center}
\includegraphics[width=8cm]{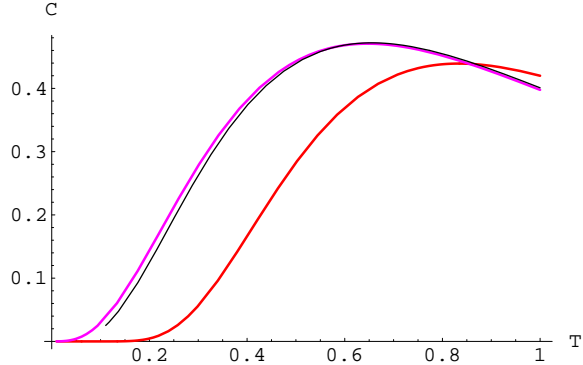}
\end{center}
\caption{\label{fig5027}Specific heat for the new mixing entropy ($\epsilon$=0.1, black line), Tsallis ($\epsilon$=0.2, pink line) and B-G (red line)}.
\end{figure}

%%%%%%%%%%%%%%%%%%%%%%%%%%%%%%%%%%%%%%%%%%%%%%%%%%%%%%%%%%%%%%%%%%%%%%%%%%%%%%%%%%%%%%%%%%%%%%%%%%%%%%%%%%%%%%%%%%%%%%%

%%%%%%%%%%%%%%%%%%%%%%%%%%%%%%%%%%%%%%%%%%%%%%%%%%%%%%%%%%%%%%%%%%%%%%%%%%%%%%%%%%%%%%%%%%%%%%%%%%%%%%%%%%%%%%%%%%%%%%%

%%%%%%%%%%%%%%%%%%%%%%%%%%%%%%%%%%%%%%%%%%%%%%%%%%%%%%%%%%%%%%%%%%%%%%%%%%%%%%%%%%%%%%%%%%%%%%%%%%%%%%%%%%%%%%%%%%%%%%%

%%%%%%%%%%%%%%%%%%%%%%%%%%%%%%%%%%%%%%%%%%%%%%%%%%%%%%%%%%%%%%%%%%%%%%%%%%%%%%%%%%%%%%%%%%%%%%%%%%%%%%%%%%%%%%%%%%%%%%%

%%%%%%%%%%%%%%%%%%%%%%%%%%%%%%%%%%%%%%%%%%%%%%%%%%%%%%%%%%%%%%%%%%%%%%%%%%%%%%%%%%%%%%%%%%%%%%%%%%%%%%%%%%%%%%%%%%%%%%%

%%%%%%%%%%%%%%%%%%%%%%%%%%%%%%%%%%%%%%%%%%%%%%%%%%%%%%%%%%%%%%%%%%%%%%%%%%%%%%%%%%%%%%%%%%%%%%%%%%%%%%%%%%%%%%%%%%%%%%%

%%%%%%%%%%%%%%%%%%%%%%%%%%%%%%%%%%%%%%%%%%%%%%%%%%%%%%%%%%%%%%%%%%%%%%%%%%%%%%%%%%%%%%%%%%%%%%%%%%%%%%%%%%%%%%%%%%%%%%%

\section{Maximum Mixing and Broken Symmetry}

In the case of Boltzmann-Gibbs  or Shannon entropy, the maximum entropy state corresponds to the symmetric state when the probabilities associated with the states are equal.

It could be expected that the maximum mixing entropy would also pertain to the most symmetric case when all states exist in equal proportions, so that there is enough diversification in the state space to produce mixture and  uncertainty.

However, a simple calculation with three states shows a very different picture when $q>1$.

In view of symmetry, the intuitively maximal mixing entropy value should be given by
\begin{equation}\label{ssb1}
S_{max}= - 3 p^q \log p
\end{equation}
when $p=1/3$.

If the constraint $p_1+p_2+p_3=1$ is imposed, the entropy is given by

\begin{equation}\label{ssb2}
S= - p_1^q \log p_1 - p_2^q \log p_2 - (1-p_1-p_2)^q \log(1-p_1-p_2)
\end{equation}

\begin{figure}[ht!]
\begin{center}
\includegraphics [width=8cm]{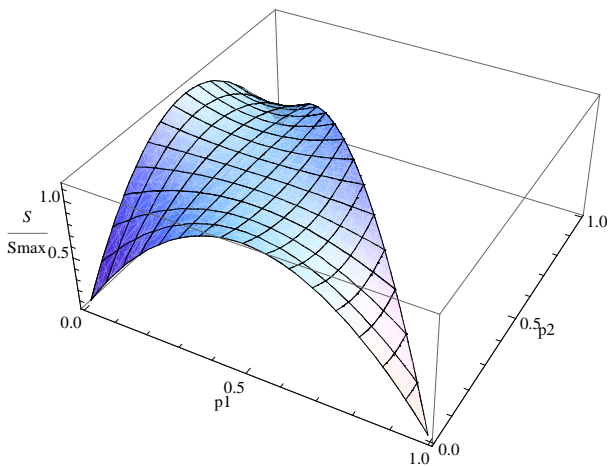}
\end{center}
\caption{\label{figssb1} Our entropy for a three-state system, with parameter $q=2.44$, as two independent
probabilities $p_1$ and $p_2$ are varied wit the constraint $p_1+p_2+p_3=1$. The expected maximum at the symmetry
point $p_1=p_2=p_3$ turns out to be a local maximum. The global maxima are not at the end points with one of the
probabilities going up to unity and the others vanishing, which gives zero entropy as expected, but occurs near
such end points, as shown clearly in the next figure.}
\end{figure}

\begin{figure}[ht!]
\begin{center}
\includegraphics [width=8cm]{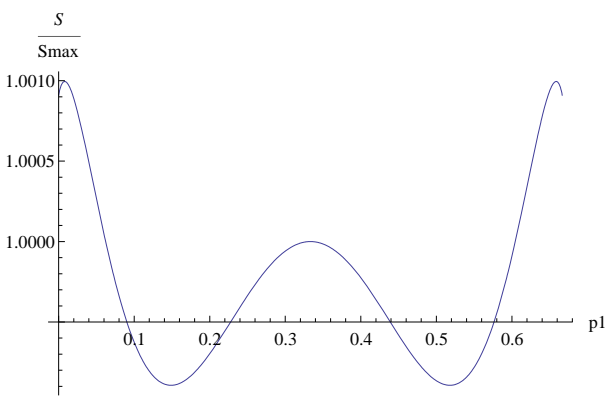}
\end{center}
\caption{\label{figssb2}Two-dimensional version of the previous Fig. ~\ref{figssb1}, with $p_1=1/3$ fixed, so that only
$p_2$ varies. This shows a clearer picture of local maximum at the symmetry point and global maxima near the end
points.}
\end{figure}

 A plot of $S/S_{max}$ in Figs.~\ref{figssb1},~\ref{figssb2} for $q=2.44$ shows a very interesting departure from the expected maximum. Although the point of symmetry $p_1=p_2=p_3=1/3$ shows a local maximum, global maxima are seen very close to the points where one of the states has a probability close to one, but the others have very small probabilities.  The end points, where one state has probability 1, still show zero entropy.  The plot appears very similar to that of spontaneous symmetry breaking in field theory.

If $q\leq 1$ the maximum entropy is given by the expected point of symmetry.

 When a posterior PDF is constructed using Bayesian rules from a uniform (Dirichlet \cite{dirich}) prior PDF and a very scarce set of data-points while maintaining the total probability summation constraint, it can again be seen that the expected value of the mixing entropy  shows a peak for asymmetric $p_i$ values, exceeding that at the symmetry point.  \cite{WB1}.  Details of the broken symmetry with respect to the new mixing entropy and data can be found in \cite{FS2}

\subsection{Origin of the Broken Symmetry}

%%rewrite with math
The origin of the broken symmetry can be understood when mixing is seen with respect to the total number of mixed states.  In the asymmetric case, a few states dominate, while the probabilities for the other states are low.  When $q >1$, the probability of one of these many rare states remaining pure becomes  even smaller very quickly while only a few high probability states have a chance of remaining pure.  When the probabilities are symmetric, the equally expected states have a similar chance of getting mixed, and the effect of $q$ is equally distributed among the states so that the chance of each of the states being observed in a pure state is still large.

\section{Quantum and Classical Stochasticity}
The paradigm of quantum measurement and the interplay between quantum and classical information remains a fundamental problem in physics. Several models have been proposed \cite{Zurek,Sewell}.  In related papers, \cite{FSWALK,FSIMAGE}, the expression of a quantum state within a macroscopic detector state has been discussed. Here, we specifically present \cite{FS3} the emergence of classical stochasticity from  a quantum picture and the level of uncertainty from measurement and approximations of quantum information related to a scaling factor, $q$, introducing distortion in the phase space because of interactions.

A quantum version of the new mixing entropy would measure the uncertainty related to the approximation in the  expression of quantum states within the classical domain.

We begin by reviewing the basic notions of entanglement and stochasticity, and how stochasticity can be an expression of the degree of entanglement of a system with the encompassing environment.

\subsection{Measures of Stochasticity}

A density matrix of the form

\begin{equation} \label{rhogam}
\rho = \left[ \begin{array}{ccc}
c^2 & \gamma c s \\
\gamma c s  &    s^2\\
\end{array} \right]
\end{equation}
represents an entangled state of two qubits.

$\gamma=1$ gives the pure quantum (entangled) state

\begin{equation} \label{psient}
|\psi\rangle  = c |0\rangle |0\rangle +  s |1\rangle |1\rangle
\end{equation}

In the above equations $c= \cos(\theta)$, and $s=
\sin(\theta)$. The entanglement is with respect to the basis vectors
$|00\rangle$ and $|11\rangle$.

 $|\gamma | <1$, represents an impure state with classical stochasticity. Maximum stochasticity is given by $\gamma=0$.

\subsection{Stochasticity from Entanglement with Environment}

If the environment(E)-quantum (AB) system can be seen as an entangled pure state:
\begin{equation}\label{PsiABE}
|\Psi_{ABE}\rangle = \sum_{ijk} c_{ijk} |i\rangle_A |j\rangle_B
|k\rangle_E
\end{equation}
with a density matrix

\begin{equation}\label{rhoABE}
\rho_{ABE} = \sum_{ijk,lmn}c_{ijk}c^*_{lmn}|ijk\rangle\langle lmn|
\end{equation}
the stochasticity of the quantum pair can be found by tracing the environment out from the pure state:

\begin{equation}\label{rhoABcc}
\rho_{AB} = \sum_{ij,lm}c_{ijk}c^*_{lmk} |ij\rangle\langle lm|
\end{equation}

If the entangled pair comes as $|00\rangle$ and $|11\rangle$, two angles $\theta$ and $\theta'$  can be defined such that

\begin{eqnarray}\label{carr}
c_{000} = c c'  \nonumber  \\
 c_{001} = c s'  \nonumber  \\
  c_{110}= s s' \nonumber  \\
c_{111}= s c'
\end{eqnarray}

Tracing out $\it{H_E}$ gives

\begin{equation}\label{rhoc'}
 \rho_{AB}= \left[ \begin{array}{cc}
c^2 &  2 c s c's'  \\
2 c's'c s  &    s^2
\end{array} \right]
\end{equation}

This can be compared with the original density matrix for the entangled pair given in the last section to see that

\begin{equation}\label{gamma}
\gamma = \sin(2 \theta')
\end{equation}

Hence, stochasticity can be found from the angle of coupling of the quantum system with the environment.
%%%%%%%%%%%%%%%%%%%%%%%%%%%%%%%%%%%%%%%%%%%%%%%%%%%%%%%%%%%%%%%%%%%%%%%%%%%%%%%%%%%%%%%%%%%%%%%%%%%%%%%%%%%%%%%%%%%%%%%%%%%%

\subsection{Entanglement with the Environment and Preferred Bases}
\cite{Zurek} has described the possibilities of preferred environment states corresponding to quantum states. In the density matrix in the previous subsection, $c'= \cos(\theta')$ and $s' = \sin(\theta')$.  These angles are defined such that the cosine part gives a measure of the system state entangled with the same preferred state of the environment.  These angles, therefore, are measures of the existence of a preferred environment basis.  In the special case of $\theta'= \pi/4$ there is no preferred basis, since similar and dissimilar states of the environment and the system are seen entangled with the same probabilities, and, so, no quantum measurement or collapse takes place within the environment, and the system exists in a pure entangled wave with no stochastic component within the density matrix.

%%%%%%%%%%%%%%%%%%%%%%%%%%%%%%%%%%%%%%%%%%%%%%%%%%%%%%%%%%%%%%%%%%%%%%%%%%%%%%%%%%%%%%%%%%%%%%%%%%%%%%%%%%%%%%%%%

\section{Entanglement, Entropy and Mutual Information}

Mutual information between two subsystems is a measure of the information common to both the subsystems, and gives a measure of the correlations between the subsystems.  In terms of Von Neumann entropy, the mutual information of two subsystems measures the difference between the summed entropies of the individual subsystem and the entropy of the combined system. It is, therefore, a measure of uncertainty of the combined system arising from the correlations between the subsystems.

This can be written as

\begin{equation}
I_{AB} = S_A + S_B - S_{AB}
\end{equation}

Similarly, if an entropy is a measure of uncertainty arising from deformation of phase space of individual components of a system, the mutual information relates to the origin of the uncertainty by looking into the uncertainty of the combined system, and that of individual subcomponents when measured separately. The new mixing entropy is a measure of such uncertainty, and the same mutual information relation as that for Von Neumann entropy holds, after substituting the Von Neumann entropy with the new mixing entropy.

If the entropy of a system involves the component $\log \rho$, it has a value of zero for the case of pure systems, since for a pure system, one of the eigenvalues of $\rho$ is one and the rest are zeros. This obviously indicates that a pure wave function has no mixing with other systems.

In the case a single quantum system is interacting with the environment, we have,

\begin{equation}
I_{AE} = S_A + S_E - S_{AE}
\end{equation}

By Araki-Lieb condition, \cite{AR1},

\begin{equation} \label{Araki}
S_{AE} \geq |S_A - S_E|
\end{equation}
$S_{AE} = 0$, because the combined system is pure.  So, $S_A =S_E$.

\begin{equation} \label{Armut}
I_{AE} = 2 S_A \nonumber  \\
\end{equation}
For Von Neumann entropy, this gives
\begin{equation}
 = -2 Tr_A [\rho_A\log(\rho_A)]
\end{equation}

Using mixing entropy, we obtain

\begin{eqnarray} \label{myinfo}
I_{AE} = - 2 Tr_A[ \rho_A^q \log(\rho_A)] \nonumber  \\
 = - 2c'^{2q}\log(c'^{2q}) -2 s'^{2q} \log(s'^{2q})
\end{eqnarray}

%%%%%%%%%%%%%%%%%%%%%%%%%%%%%%%%%%%%%%%%%%%%%%%%%%%%%%%%%%%%%%%%%%%%%%%%%%%%%%%%%%%%%%%%%%%%%%%%%%%%%%%%%%%%%%%%%%%%

%%%%%%%%%%%%%%%%%%%%%%%%%%%%%%%%%%%%%%%%%%%%%%%%%%%%%%%%%%%%%%%%%%%%%%%%%%%%%%%%%%%%%%%%%%%%%%%%%%%%%%%%%%%%%%%%%%%%%%
%%%% start here

We now discuss a special case when a quantum mesoscopic system can be split into two entangled segments $A$ and $B$, coupled with a common environment.  Scaling and quantum correlations both become important in a meso-system, and  make it an ideal choice for studying semi-classical mixing.  This study is in the spirit of the analysis carried out by \cite{JKCH} regarding Von Neumann and Renyi entropies of spin chains placed in a common field.

Once again, we assume that the entire system comprising of all the components and the environment is a pure quantum wave so that the components arise as local approximations of complex wave functions.

\begin{eqnarray}
I_{ABE}(q) = -S_{ABE}(q)+ S_{AB}(q)+S_{BE}(q) + S_{AE}(q) \nonumber
\\ -S_A(q)-S_B(q)-S_E(q)
\end{eqnarray}

$S_{ABE}(q) = 0$  for all values of $q$.

The density matrix for the subsystem $A-E$ can be obtained by tracing over $B$:

\begin{equation}\label{rhoAE}
 \rho_{AE}= \left[ \begin{array}{cccc}
c^2c'^2 & c^2c's' & 0 & 0\\
 c^2c's' & c^2s'^2 & 0 &0\\
 0 & 0 & s^2s'^2 &  s^2c's' \\
 0  & 0 & s^2c's'
& s^2c'^2
\end{array} \right]
\end{equation}.
A similar form can be obtained for $\rho_{BE}$.

Using the eigenvalues of $\rho$, the mutual information becomes

\begin{eqnarray}  \label{3mut}
I_{ABE}(q) =  c'^{2q} \log(c'^{2q})+ s'^{2q} \log(s'^{2q})
\nonumber \\
 -\lambda_+^{q} \log (\lambda_+^{q})-
\lambda_-^{q} \log (\lambda_-^{q})
\end{eqnarray}
%%%%%%%%%%%%%%%%%%%%%%%%%%%% stpped here
Here,
\begin{equation}
\lambda_{+,-}= (1/2) ( 1 \pm {\surd[{1- 4( 1- \gamma^2) c^2s^2}}])
\end{equation}

and $\gamma$ =$\sin 2 \theta \prime$.

If $1-\gamma$ represented the stochasticity in an $A-B$ system after the environment is traced out, the mutual information of the $A-B$ part is

\begin{eqnarray}\label{ABmut}
I_{AB}(q)  = -S_{AB}(q)+ S_A(q) + S_B(q) \nonumber  \\
 =\lambda_+^{q} \log(\lambda_+^{q})+ \lambda_-^{q} \log (\lambda_-^{q})
 \nonumber \\
 - 2 c^{2q}\log(c^{2q}) - 2 s^{2q} \log(s^{2q})
\end{eqnarray}

\begin{figure}[ht!]
\begin{center}
\includegraphics[width=8cm]{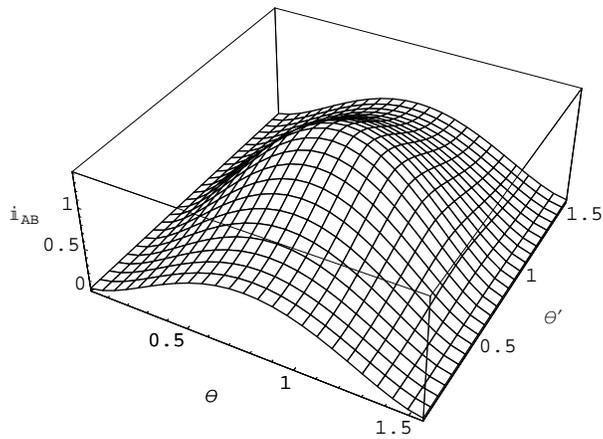}
\caption{\label{fig6036}$I_{AB}$ as a function of entanglement angle
$\theta$ in $A-B$ space and the entanglement angle $\theta'$ with the
environment, which is related to the stochasticity.}
\end{center}
\end{figure}

Fig.~\ref{fig6036} plots the mutual information of this system as a function of $\theta$ and $\theta'$ when Von Neumann entropy is used.  This also corresponds the $q=1$ case for the new mixing entropy.

%%%%%%%%%%%%%%%%%%%%%%%%%%start from here

\begin{figure}[ht!]
\begin{center}
\includegraphics[width=8cm]{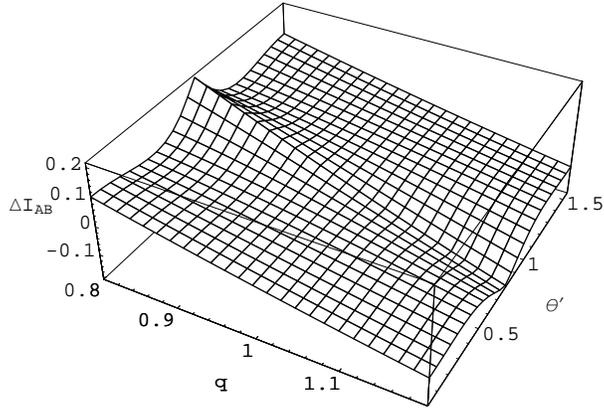}
\caption{\label{fig6037}Difference of MI from our entropy with that
from Von Neumann entropy; $\theta=\pi/4.$}
\end{center}
\end{figure}

\begin{figure}[ht!]
\begin{center}
\includegraphics[width=8cm]{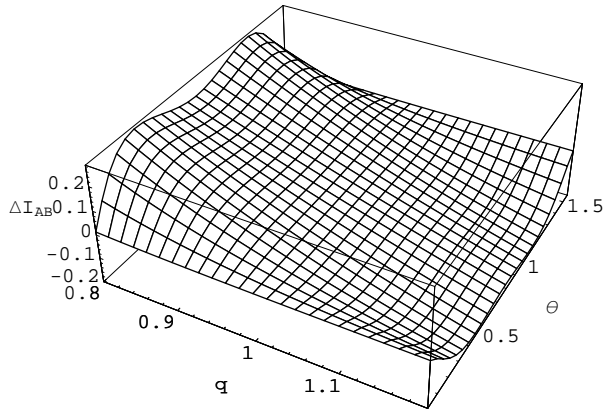}
\caption{\label{fig6038}Similar to Fig.~\ref{fig6037}; $\theta'=\pi/4$.}
\end{center}
\end{figure}

$\Delta I_{AB}$  is a measure of the difference of the mixing entropy and the Von-Neumann form in mutual information.  The effect of  $q$ with $\theta'$ and $\theta$ are shown in in Fig. ~\ref{fig6037} and Fig. ~\ref{fig6038} respectively.
In each case, one angle is varied while the other is kept fixed at $\pi/4$. A symmetry can be seen around $\pi/4$. The role of the environment is pronounced, and can be seen as a sharp peak or dip near $\theta'= \pi/4$, which corresponds to no stochasticity. The mutual information between the subsystems of $A-B$ derive entirely from the correlations between $A$ and $B$, and is, hence, highly sensitive to the distortion of the phase space of $A-B$.

%%%%%%%%%%%%%%%%%%%%%%%%%%%%%%%%%%%%%%%%%%%%%%%%%%%%%%%%%%%%%%%%%%%%%%%%%%%%%%%%%%%%%%%%%%%%%%%%%%%%%%%%%%%%%%%%%%%%%%

%%%%%%%%%%%%%%%%%%%%%%%%%%%%%%%%%%%%%%%%%%%%%%%%%%%%%%%%%%%%%%%%%%%%%%%%%%%%%%%%%%%%%%%%%%%%%%%%%%%%%%%%%%%%%%

Fig.~\ref{fig6037} plots the effect of $\theta'$ for
varying $q$. Fig.~\ref{fig6038} shows the region
( $q < 1 $) for different $\theta$ values.

\begin{figure}[ht!]
\begin{center}
\includegraphics[width=8cm]{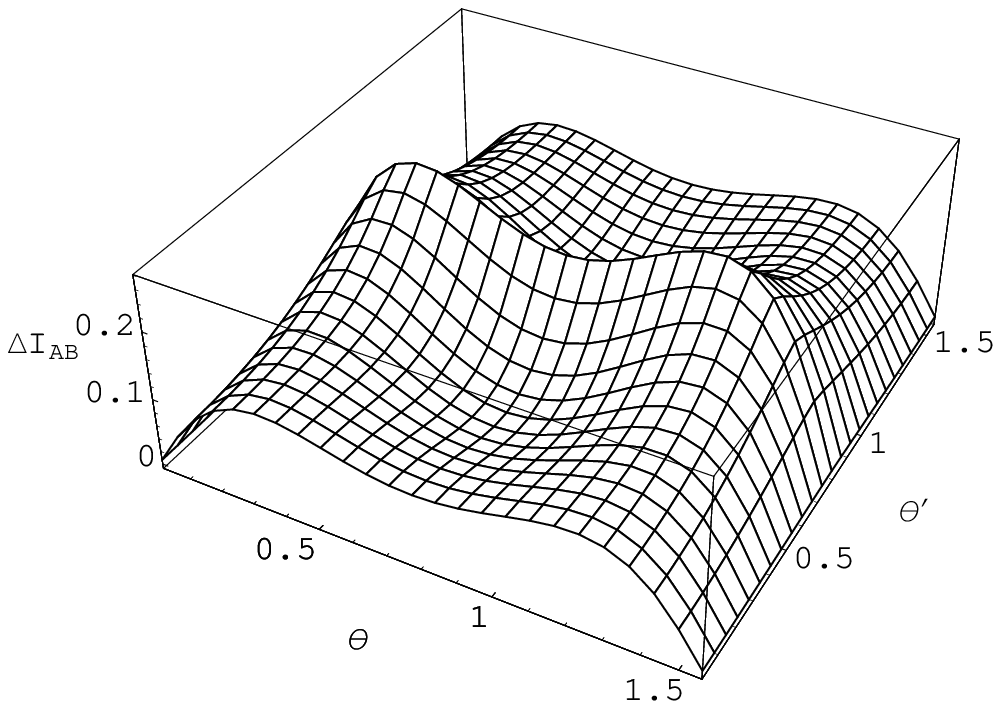}
\caption{\label{fig6039}MI difference between our entropy form and
Von Neumann for $q=0.7$.}
\end{center}
\end{figure}

\begin{figure}[ht!]
\begin{center}
\includegraphics[width=8cm]{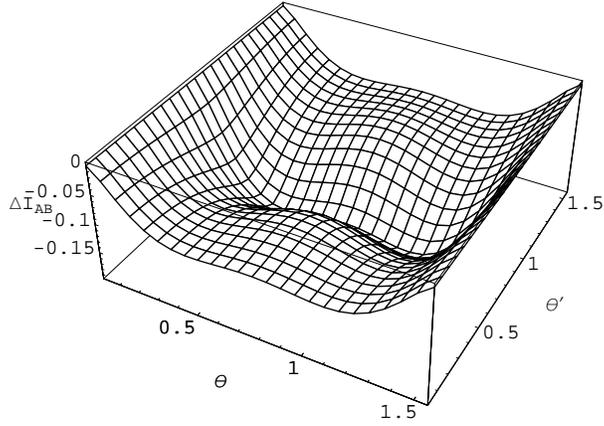}
\caption{\label{fig6040}Same as Fig. ~\ref{fig6039} but for $q=1.3$.}
\end{center}
\end{figure}

\section{Origin and Philosophical Implications of the New Entropy}
In this section, the origin of the new form of mixing entropy is discussed in detail.  The implications of the terms such as cell, deformation and letters are also analyzed in the context of physical situations.  The relations of the new mixing entropy with the basic ideas of organization and emergence of information and identities are also elaborated.

\subsection{Measurement of Information and Meaningfulness}
When a measurement is able to yield a pattern which is predefined and meaningful, it identifies a recognizable state.  The set of such patterns comprise an alphabet in our model.  Each predefined pattern is a letter.

The notion of meaningfulness comes from piling or organization of information.
When a certain pattern or letter is able to associate with a group of correlated ideas, and hence define certain properties or characteristics, information is conveyed. Hence, a system is informative only when a set of information can be isolated that is recognizable because of previous experience.

A measurement necessitates the measurer or observer to interact with the system at a certain scale.  Hence, the measurement yields a set of information related to the property of the measured system defined at that scale.  These properties originate because of the organization and scale dependent correlation of the system.  One such measurement  yields a unit pattern connected with a defined set of information. When the information containing cells, indicating the scale of a unit, are robust, information of one cell is contained entirely within that specific cell. Separated piles of information can be protected within each such cell.  So the identity of a cell is defined by the localized correlations of characteristics and, hence information.  The degree of permeability of the cell and the maximum phase space permitted prevents information of one cell from leaking into another cell.

A very simple example of cells can be found in combinatorics, when a ball is placed in a box, and the box may contain a red ball or a black ball.  Here, the cell would be a box, which can contain only one ball, and the letter would be red or black, which would define the color of the ball and hence convey a certain group of information about the state of that cell.

\subsection{Meaningfulness of States and Cells}
The amount of information carried by the cell is dependent on its volume, and the information phase space accommodated by it. While it is possible for the cell to contain any information, the amount restriction makes it necessary to choose subsets of possible information and exclude the rest. This breaking of symmetry by the acquisition of specific information \cite{collier} associates characteristics with the cell.  Cells acquiring different subsets of information may exhibit different characteristics and hence possibly distinct states. The constraint of information content makes these cells appear in one of many possible states at a time. While an elementary particle may be a cell for quantum states, an aggregation of particles may also be a cell, the states indicating the many possible characteristics of the cluster.  As systems become complex, the cells accommodate diverse components, and the states of the system depend on the interactions among these components. Again, while the cells of elementary particles, such as electrons, are well-defined, cells in more complex systems may be more loosely defined, exhibiting varied degrees of stabilities.  A country, acting as a social cell, for example, does not have a defined population. All countries, however, would still display possible states of certain characteristics.

\subsection{Disorder in Mixing of States}
The mixing of states may be important if we consider how objects are identified within an environment, and are expressed as unit objects.
Categorization of objects makes it possible to define macro-states that are expected to occur repeatedly in the environmental soup. For examples there are many ``trees'' in the environment.  Again, within trees, there are many palm trees that repeat themselves in the environment.

Since such a state only depends on averaged or summed macro-properties, it can be a pure state even when the constituent microstates are defined only within a range. A probability associated with such a pure state indicates the possible frequency of the emergence of the same state.  The state repeats when the strongly correlated characteristics of the state can emerge within the environment, with macro-properties within the given bounds. It is initiated by repeated conditions or blueprints and static physical laws. These states stabilize if the information within them, and hence their characteristics, are well separated from the characteristics of other possible states.  When such states are mixed, the characteristics may become uncorrelated, and the set of characteristics in the form of a mixed state may not be repeated as easily or at all, as each mixing may be uniquely defined, making the probability distribution of discretely defined states previously summing to unity erratic.

A degree of order or robustness inherent within an ensemble of information-carrying cells is expressed by its ability to retain pure known patterns, that are categorized by means of separable traits and predefined knowledge even if the phase space is slightly distorted.  The possibility of emergence of unknown characteristics and hence new degrees of freedom and a larger number of identities with a slight distortion of these cells may indicate a new type of disorder characteristic of the system. This may define a new type of entropy directly related to the basic principles of information and identity.

We take these ideas into account to construct a new mixing entropy based on the identification of recognizable letters within an ensemble or a sequence.

\subsection{Information Mixing and Uncertainty of Identities In Complex Systems}

 Let a single cell of size unity contain a complete set of information required to assign the entire cell a predefined letter. A deformation of the cell's information containing phase space would permit either more or less information within the register.  As a result, partial letters or subsets of information previously contained in other cells or in the environment would be allowed to leak into the deformed cell or flow out of it.  Since the entire cell is expressed as a unit by a measurement, this deformation may cause the identity of the letter contained within the cell to change, allowing fractional letters to be accommodated in the form of leaked incomplete information.

This is possible only when the cells are not completely closed, but semi-closed \cite{FS8} so that information within the cell itself is strongly correlated (and causes the identity of the cell to change little when the cell is not deformed and is ''filled" with strongly correlated information) and expressed as a predefined identity or letter. If the identity of the cell is also loosely related to its environment and other cells (i.e., the cell wall is semi-permeable), a deformation of the cell volume may cause leakage of information.  After information is leaked in or out, the renewed components within the cell may become strongly correlated again within the cell, redefining the identity.

Often the unit in a complex system is made of components and the property of the system reflects its constituents. For example, in neural systems, neurons are interconnected in such a manner that outputs from each neuron, using simple rules depending on input signals, sum up to produce complicated neural signal patterns displayed by the brain.  These trains of signals within a unit brain make up the reactions and personality traits of an individual.  Adding or subtracting a few neurons may change the pattern exhibited by the individual. Causing one person to take into account another person's experience and act upon another person's experience, hence distorting the ''neuron phase space" of an individual's brain may cause different types of reactions or personalities to be expressed together by an individual.

Similarly, a bacterial colony may be comprised of individual bacteria.  If one bacterium is allowed to breed to form a colony of an optimal size, each constituent of the colony will be a bacterium of the same strain. If the colony itself is taken as an identity unit, the characteristic of the colony would reflect only the characteristics of one strain of bacteria. If separate bacterial colonies exist consisting of different strains of bacteria, a measurement or detection of a colony will yield the characteristics of the
bacteria associated with the colony.  Such colonies may be represented by pure letters indicative of the single strain of bacteria associated with the colony. However, if the optimal colony size is increased because of any change in environmental conditions, migration or accommodation of a different strain by spontaneous growth may cause the colonies to contain mixed characteristics. In the simplest case, this scenario would imply the creation of a statistical ensemble with a combination of different strains instead of one within each colony, giving a new statistical ensemble for a colony.  However, in a realistic scenario, interactions among separate strains and expression of dominant characters and suppression of certain characteristics because of competition may also come into being within each colony because of the mixing.  This is important when the entire colony is expressed by a measurement, and is taken to be a letter.

Breeding between species of bacteria within a single newly defined colony may create novel forms of bacteria with new characteristics within the colony as well. Although this again would imply reshuffling the genetic material of each individual bacterium, it is not each of the genes that is expressed separately.  An entire bacterium is expressed as a entity with bundled properties: the environment (existence of other strains and other environmental factors), sequencing of genes, and interaction among parts of the genetic material within the bacterium unit help create the identity of the new strain of bacteria, facilitating certain behaviors while suppressing others.  Some of these newly expressed characteristics may not have been present in any of the pure strains before. This situation is similar to deforming a cell (which is a colony) and allowing letters (indicating strains) to mix.

In terms of cultural cells, superposing modes or traits of foreign cultures on top of one consistent set of logically coherent traits of a single culture, and assimilating these imported traits into the characteristics of a population may be analogous to deforming the phase space of social norms of a previously unit identity.

\subsection{Measurement and Mixing of Information}
Mixing of letters may come into being from the physical deformation of information containing systems as explained before.  However, similar scenarios may arise because of deformation of the measurement phase space since the expression of an identity is related to the interaction between both an object and a measurer.

The phase space of measurement and the measured cell may not be the same due to imprecision and limitations of measurement.  The measurement of a minuscule state may involve the interaction of a measuring device and a measured system at a mismatched scale so that information contained within one cell is not retrieved.  Hence a measurement may be contaminated, including information from neighboring cells.

Let us assume that five different types of soil samples can be categorized based on their components and the ratios of the components (such as minerals and vegetation). Each sample of soil would have the components distributed within each cell or sample patch.  A unit cell then would contain one of five possible soil types.  Now, if a measuring device collects samples from a cell to identify a soil type, then an imprecision of the measuring device's resolution may cause soil types to get mixed in a collected sample.  This mixture would cause vegetation and minerals of different soil types to interact within the collected sample.  The leakage of information between ''pure letter" samples because of such deformation of cells caused by the interaction between the measurer and the system may also be a source of mixing.

The mixture would increase the degree of freedom in the state space, since new identities will be created due to leakage of information from one system into another

\subsection{Nonextensivity from Leakage}
In a closed system, information is contained within the system.  However, when a system is semi-closed \cite{FS8}, information can be exchanged with the environment to some extent, as described before. Deforming the scale of each cell in such a system may thus cause information leakage, making the system to non-extensive.

Let us consider the simple example of bacterial colonies once again.
At the most elementary level, the environment has its elements correlated and organized into infinitely many possible expressible ''objects" of various sizes and degrees of complexity.  Some of these objects may be bacteria.  It is possible to identify a subsystem consisting of bacterial colonies, and consider the rest as the environment. The colonies containing different pure strains (at optimal colony sizes) are considered to be pure letters in a cell.  However, new bacteria can be created by using environmental elements (eg. floating spores of bacteria), adding to the members of a colony, and bacteria may die, becoming part of the residual environment.  Hence, the system of bacteria is not closed, but may be considered semi-closed.

The deformation of bacterial colonies and the mixture of strains were discussed previously. The addition of new mixed identities cause the total possible measurable identities to increase, and the identities of the new mixed types would now depend on the probabilistic mixture and the possibility of introducing a bacterium of a different strain spontaneously from the environment if the colony size is suddenly increased.

Hence, the previously ordered partitioning of information into different strains, creating defined organizational structure in a complex ecological system now allows the piles of separated qualities to mix, increasing the disorder in categorization.

The nonextensivity of the entropy measuring the disorder depends on the exchange of information among semi-closed cells in contact with the environment. Increasing the size of each cell of a previous collection of a fixed number of cells changes the entire system to a new deformed system by deforming the phase space of each information containing cells, and hence the entire system of cells.  So the collection of cells has a new amount of partitioned expressible information distributed among themselves, and not just the previously identified letters re-partitioned into phrases of different sizes adding up to the exact same information as before.

The nonextensivity here derives from the total possible states changing because of the scale associated with the states and the emergence of new states at that scale.  New states yield new possible information depending on the correlation among the subsystems that are expressed.  Since the leakage of information depends on interactions among the predefined cells and also with the environment, the total disorder is not a sum of the individual disorders of subsystems.

In the case of partitioning a previously known sequence, such as a DNA sequence, changing the length of the partitioning n-tuple would imply covering the same sequence with tiles of a different length, preserving the total content of the sequence. So the sequence itself does not undergo any physical changes.  However, in this case of deformation and mixing, the system itself is deformed because additional material can flow in from or out to the environment, or material within the cells can mix freely, while information among cells are kept more separated.

When the case of measurement is considered, so that a deformed cell represents an imprecise measurement that takes into account information from a neighboring subsystem, corrupting the measured state, the total measurement is not about re-binning the sequence with tiles of a different lengths, but allowing the same information to be detected twice, in two different cells, because of the mismatch in information containing cell dimension and measurement-cell dimension due to the imprecision of measuring device. This causes information to leak, and the measure of entropy in such systems to become non-extensive.

\section{Comparison with Other Entropy Forms}

In typical thermodynamic entropies, the number of possible distributions of microstates yielding a constant macrostate such as energy would determine the disorder of the system.  If more microstates are available to the system, a larger number of reorganizations within the system can yield the same macroscopic variable.

In the mixing entropy, the disorder originates in the possibility of losing a pure identifiable
state because of deformation of information containing cells or measurement units. Mixing causes the discrete set of letters to attain a different q-deformed distribution, allowing for new forms of identities and hence states to emerge. This entropy is the measure of disorder inherent within the system at a certain expressible scale that causes its constituents to lose their identities upon slight distortion. The state space expands by accommodating previously unknown mixed states.  Hence the mixing entropy is related to the order in the environment produced by separating classes of information by creating correlations among possibly coexisting characteristics.

The mixed letters, however, are not predictable, since the mixing process allowing
partial letters to occupy a cell is continuous while the predefined set of letters is discrete. The emergence of new characteristics also depends on the number of components contained within each letter's information pool that are
allowed to mix with another letter.

In information theory, the traditional entropies indicate the uncertainty associated with the system.  If a sequence of letters are read from the source, the entropy indicates the uncertainty related to predicting the next letter, given all measurements are expected to yield a letter from a known alphabet.

Again, in quantum systems, the traditional entropy measures the departure of the system from a pure quantum state, and hence indicates the existence of a mixed ensemble, which can be defined by a density matrix. Although the words pure and mixed are used in the quantum context, each measurement in such a quantum system is expected to yield one pure quantum state that is previously defined, and the mixture is in the existence of multiple pure states within a statistical ensemble.  The pure states themselves do not change. For example, if two pure
states $\psi$ and $\phi$ can be detected in a mixed ensemble, the density matrix is given by
\begin{equation}
\rho =p_1 |\phi><\phi| +p_2 |\psi><\psi|
\end{equation}
 Each measurement would still yield either $\phi$ or $\psi$ but with
probabilities $p_1$ and $p_2$.

The density matrix is an attempt to assign statistical properties to a quantum ensemble by associating probabilities with each possible pure wave function instead of considering the wave amplitudes.  The formulation manages to map a quantum ensemble into a Gibbs/Shannon type ensemble.  The correlations among subsystems are given by mutual information.  Thus quantum entropy may also act as a measure of quantum entanglement.  A pure quantum wave has a rank one density matrix, with an eigenvalue of one for one particular pure wave, and zero for others, making the corresponding entropy zero.  The entanglement among subsystems can be found by subtracting the entropies of the subsystems using reduced density matrices from the total entropy of a pure system, which is zero.

However, the process of measurement that expresses a quantum state in a classical world, and the effect of measurement phase spaces that define each of the systems within the ensemble are not taken into account.  An interesting case is that of a mesoscopic systems where quantum and classical properties -- both come into play.  Moreover, a mesoscopic system can be divided into subsystems that are measured independently.  When subsystems are measured, the measurement interface is important, and any distortion or imprecision of the measurement interface may cause the presence of neighboring subsystems to be recorded, corrupting data, and making detected states undefined.  If the entire mesoscopic system and its environment are taken as a pure wave,which is defined, the subsystems that are measured separately give mixed ensembles and the entanglements can be measured by tracing out the subsystems from the entire system.  However, the compartmentalization of sub-systems here is crucial, and any distortion or scaling of subsystems will redefine known properties associated with previously defined subsystems by including effects of neighboring members. Hence, any distortion of the measurement phase-space, or any change of the physical properties of the meso-system, causing the subsystems to become distorted, will redefine measured quantities that were previously assigned to each subsystem. So by taking into account meso-systems with changeable sizes, and by associating correlation and density matrices with possible expressions of a quantum state within a larger system, the effect of correlation in measurement and collapse can be observed.

\subsection{Comparison with Specific Special Entropies}

Shannon's entropy was formulated by using a set of postulates associating probabilities with information. A further generalization was obtained by allowing the probabilities to carry weight.  In Renyi entropy, a weight or continuous moment, $\alpha$, is associated with a probability distribution so that each probability $p_i$ is mapped to ${p_i}^q$.  The entropy is given by the mathematical expression:

\begin{equation}
S_R= {1/[{1-\alpha}}{log\sum_{i=1}^N P_i^{\alpha}]}
\end{equation}

Renyi entropy was constructed as the most general form of information measure that is also additive, and is hence extensive.  In his original paper \cite{RE1}, Renyi described his measure as the ''entropy of order $\alpha$" of the distribution P.

The use of the weight factors, $\alpha$, or using moments of $P_i$ instead of $P_i$ enables one to find a more comprehensive statistical distribution of the system in the same manner calculating successive moments of a variable enables one to find a more realistic function when the distribution function itself cannot be accurately calculated.

The use of moments also enables one to amplify different regions of the probability distribution. If a value of $\alpha$ is low, then regions with low probability distributions are amplified, while if $\alpha$ is large, regions with a high probability distribution are emphasized.

Moments of probability may be useful when comparing several probability distributions \cite{RE1}.

Renyi entropy is also used for detecting multifractals \cite{JIZB}, where several components make up a system and their probabilities scale with different fractal dimensions.  When all the probabilities have the same fractal dimension, the normalized moments also have the same distributions for different values of $\alpha$.  However, if several fractal distributions are intertwined within a system, the moments of the probability distribution yield different patterns.

In a multifractal, each of the probability components is intertwined with the other components and scale differently.  So detecting a cell or box yields several components interlaced, and rescaling the box would again yield all of the components, but scaling differently, each as
\begin{equation}
P_i = L^{\alpha_i}
\end{equation}
Here, $\alpha_i$ is the Lipschitz-Holder exponent, which is connected to the fractal dimension $f(\alpha_i)$.

 In a distribution with various cell sizes, a multifractal distribution would cause the components to display power-law behaviors with different exponents, and Renyi entropy can be applied to study such distribution.  One example is in the case of multifractal distribution of soil particles \cite{SO1}.

While Renyi entropy is also used in regard with sequences where identities are expressed at different scales, the partitions completely cover the space, and the distribution measure is still extensive.  One such case is the analysis of DNA sequencing \cite{DNA}. In DNA sequencing, entropy can be measured for the distribution of single letters, A, T, C and G, each representing a nucleotide base, or for N tuples, representing segments that are often repeated.  The repetition frequency of a gene dictates the expression of a phenotype \cite{DNA}. Hence, the distribution of gene segments that convey a meaning is also interesting to analyze. However, the application of Renyi entropy here is for detecting highly repetitive sequences, while making the role of junk sequences or noise low by assigning moments to the probability distribution.

The existence of correlated sequences, expressed as functioning gene segments, are studied by using statistical methods, and the gene lengths are fixed as N tuples \cite{DNA}.  In specific cases, density functions using correlations are calculated for each site, and the DNA sequence is partitioned into segments of various lengths successively to identify repeating patterns \cite{DNA}.  However, this scaling or partitioning is simply related to the process of tiling a known sequence with a conserved amount of information.  Changing the length of a single tile changes the sequence, and adds new bases, causing the tile not to be identified with that single gene sequence.  For example, the segment AAAAAA may be meaningful, and hence a 6-tuple containing all A's
may represent that gene and may be found with a high frequency.  However, if the DNA sequence is partitioned into 7-tuples, AAAAAAA does not represent the same gene, and is not repeated with the same frequency.  Hence, changing the scaling or partition size in a gene sequence and applying Renyi entropy simply helps identify highly repetitive N-tuples, that can be expressed as specific gene segments. When the DNA is tiled with n-tuples with 6 bases each, each such 6-tuple serves the role of a letter, and the distribution of the 6 tuple letters in the DNA can be studied.  However, tiling the DNA sequence with 7-tuples simple covers the DNA with different 7 tuple letters that appear with different statistical distribution based on a new alphabet consisting of 7-tuple letters.  Theoretically, the distribution of n-tuple letters can be studied with many other entropies, even with Boltzmann-Gibbs, by transforming the possible 1-tuple letter alphabet to a different alphabet consisting of n-tuples, each of teh n-tuples taken as a letter or possible variable.  This is a change in variables. The use of Renyi entropy simply assigns weights to the probabilities, so that sequences or letters that repeat are considered with more importance, while noisy segments that do not repeat are neglected for each N-tuple-letter.

In the case of the entropy proposed by us, the identity of a cell is the identity of a complex system where the cell contains information pertaining to that identity.  The cell is seen as a whole, and information in the cell can mix within the cell and be expressed as a single characteristic.  Deforming the cell may add new content to the cell.  However, if the same building blocks are added, the new cell's identity remains unchanged, and there is no mixing. The repetition of the same letter in a deformed cell does not add new information for systems used with our entropy, but the addition of different letters does.  Hence, the mixing of information in a cell for the case of the mixing entropy would imply the addition of new sets of information or new properties, causing the characteristics of the cell to change.  Hence, our entropy would measure the leakage of previously partitioned information leading to possible but not definite change in the information state space, where each state is a contained in a cell, and a mixture of previously separated information would create a new state.  The mixing entropy does not detect repetitive sequences of different lengths by changing partition sizes since changing the partition size would definitely change the sequence, which in our case, may still yield the same state if fractions of the same letter is introduced in a rescaled cell.  Hence the mixing entropy is not just about correlated segment, it is also about the expression of units as pure states, and the possibility of new identities emerging by mixing different pure states in a deformed phase space, and hence the definition of mixing entropy lies deeper in the origin of identities expressed as possible information states contained within a cell, and the mixture of previously partitioned information defining another identity.

In Renyi entropy, the weight $\alpha$ is not a measure of the scale of the partitions. It is simply the moment of the probability, and hence, is not to be confused with the q used in the case of the mixing entropy. q in the mixing entropy does not indicate moment, but a deformed cell volume that can contain a different amounts of information in the form of letters.

Aczel and Daroczy (A-D) \cite{AD1, AD2} have a form of entropy which is given by

\begin{equation}
S_{AD}= - \sum_i p_i^q \log p_i/\sum_i p_i^q.
\end{equation}

This entropy is formulated in the same spirit as the Renyi entropy, and hence, is not to be confused with the mixing entropy.
An extra denominator term is used in A-D entropy so that the weights of $\log p_i$ are normalized.

Wang's entropy \cite{WA1}, has the same expression as the new mixing entropy, but is constrained by the condition

\begin{equation}
\sum_i p_i^q = 1
\end{equation}

Hence, Wang's entropy is extensive while the mixing entropy is non-extensive. Wang's entropy takes into account systems where information is incomplete due to various possible factors including existing singularities, fractal nature of the system, the existence of unusable data points (eg. the case of a coin landing on its edge).  Hence, in a series of measurements, if the total probabilities of usable variable values
sum up to be greater or less than one, Wang's entropy re-scales the ''good values" by including a q exponent so that the total probabilities of the usable variables add to one, and the properties of Gibbs entropy are retrieved.  Wang's entropy is also used with interacting systems when the interactions are weak (eg. weakly coupled electrons).  A re-scaling of the probabilities in such cases translate an interactive system to the extensive Gibb's domain and retrieves the known Boltzmann/Gibbs statistics by approximating out the effects of interactions. Thus the use of Wang's entropy is similar to the scaling $p' \rightarrow p^q$ and rewriting Gibbs entropy as $p' log p'$ with a q-rescaled Boltzmann constant.

While Wang's entropy yields the thermodynamics of the regular Boltzmann/Gibbs distribution, and can be obtained by ignoring non-extensivity of systems that display Tsalli's entropy in a weak form, the new mixing entropy displays neither Boltzmann-Gibbs nor Tsallis thermodynamics properties, but has its own thermodynamic properties.

Wang's entropy approximates a system with existing incomplete information to the ideal extensive form.  However, the mixing entropy takes into account the origin of change in identifiable states because of leakage of previously correlated information as a result of cell deformation, and hence is associated with a dynamic property of a system in contact with a larger environment.
For the added property of taking leakage of information into account, the new mixing entropy is non-extensive, and is indeed a measure of the departure of a system from an i environment where identity units are fixed at every hierarchy level.  This mixing may occur due to physical reasons, and may give rise to new degrees of freedom within the system, and thus cause the system to dynamically
attain new equilibrium points.

Unlike the case of Wang's entropy, the process of mixing does not imply the translation of a complicated system to a manageable form by smoothing out inherent interactions or anomalies by rescaling probabilities. Rather, the mixing entropy involves the interaction between a system and the measurer and the effect of layers and levels of information contained within an information unit succinctly expressed as a letter.  The effect of losing identities and creating new degrees of freedom can be observed in a continuous manner by using the mixing entropy. The process of continuous evolution into new identities in systems deformed by interactions can be observed in many interactive natural systems including ecological ones.

The mixing entropy does not indicate rescaling of the approximated system of Wang's entropy back to its original non-extensive form either.  Wang's entropy loses detailed information  by means of rescaling, and allows a system to have only clean variables.  Wang's entropy also works only for weakly interacting systems where non-extensivity can be ignored by the rescaling process.  Retrieval of lost unmeasurable variables is not possible by rescaling since it is not possible to take a truncated series and rescale it to form a more accurate one.

Rather, the mixing entropy is indicative of a very specific physical process inherent in the basic ideas of information identification and meaningfulness of data within complex systems.  Order is seen in clusterized information and deformation of the cells that hold the information (which is the deformation of cells holding letters) causes this order to be disrupted and new possibilities, meanings or mixtures of meaning can thus be formed.  New mixed states create new correlations among information segments, and necessitate more detailed interactions for the identification, classification and use of such states.

\section{Conclusion}

%new
We have reviewed a new type of entropy resulting from mixing of information between previously separated states originating from the deformation of information-containing registers when leakage of information is allowed between cells. The derivation of the entropy from the first principles related to complex physical phenomena make it a possible candidate for studying complex systems and the evolution of separate classes or categories in the physical world. The entropy has been presented in both classical and quantum contexts and comparisons.  The emergence of behavior resembling field theoretic spontaneous symmetry breaking, because of the  new degrees of freedom introduced by mixing, presents an interesting area for further investigation.  Whether such phenomena can be observed in other forms of generalized entropies is also a possibly interesting topic. Since traditional entropies consider systems where the states defining a probability distribution are kept fixed, they are unable to define disorders emerging in complex phenomena where the states are updated because of information leakage and re-stacking.  The refinement of correlations among information pieces give rise to an interesting type of ordering within the environment.  The application of the proposed new type of entropy in practical contexts within information theory may be interesting to investigate. Physical systems where mixing in various hierarchical levels and the emergence of identities are prominent may be worth serious consideration in light of the mixing entropy.

%%%%%%%%%%%%%%%%%%%%%%%%%%%%%%%%%%%%%%%%%%%%%%%%%%%%%%%%%%%%
\section*{Acknowledgements}
The author would like to thank Prof. Philip Broadbridge for encouragement and Prof. Peter Harremoes and Prof Shu-Kun Lin for their interest and their comments that helped improve the presentation and clarity of the paper.  She would also like to thank Dr. Ignacio Sola for reading the draft, Dr. Giorgio Kaniadakis and Dr. J F Collet for sending comments on some of the original reviewed papers, and Prof. Christian Beck for reading the first reviewed paper.  She would also like to thank everyone else who read the paper to help her proofread the draft.

%==========================================================

%==========================================================
% Back Matter (References and Notes)
%----------------------------------------------------------
% Style and layout of he references

%%%%%%%%%\makeatletter
%%%%%%%%\renewcommand\@biblabel[1]{#1. }
%%%%%%%%%\makeatother
%----------------------------------------------------------
% Use the following option to include external BibTeX files:
%\bibliography{}
%----------------------------------------------------------

%==========================================================

%==========================================================
% Back Mater (Copyright and Licensing Info)
%----------------------------------------------------------

%==========================================================

\end{document}